\documentclass[pra,twocolumn,showpacs,showkeys,amsmath,amssymb]{revtex4}

\usepackage{hyperref}% hyperlinks
\usepackage{graphicx}% Include figure files
\usepackage{bm}% bold math
\usepackage[utf8]{inputenc}

\begin{document}

\title{Hamiltonian Design in Atom-Light Interactions with Rubidium Ensembles:\\A Quantum Information Toolbox}

\author{S. R. de Echaniz}
    \email{sebastian.echaniz@icfo.es}
\author{M. Koschorreck}
\author{M. Napolitano}
\author{M. Kubasik}
\author{M. W. Mitchell}
    \homepage{http://users.icfo.es/Morgan.Mitchell/MMGroup/}
\affiliation{ICFO - Institut de Ciencies Fotoniques, Mediterranean Technology Park, 08860 Castelldefels (Barcelona), Spain}

\date{\today}

\begin{abstract}
    We study the coupling between collective variables of atomic spin and light polarization in an ensemble of cold $^{87}$Rb probed with polarized light. The effects of multiple hyperfine levels manifest themselves as a rank-2 tensor polarizability, whose irreducible components can be selected by means of probe detuning. The D$_1$ and D$_2$ lines of Rb are explored and we identify different detunings which lead to Hamiltonians with different symmetries for rotations. As possible applications of these Hamiltonians, we describe schemes for spin squeezing, quantum cloning, quantum memory, and measuring atom number.
\end{abstract}



\maketitle

%***************************************
\section{\label{sec:intro}Introduction}
%***************************************

The pursuit of quantum networks has recently led to much interest in developing an efficient quantum interface between atoms and light, where the first are a good medium to store and process information, and the latter is a good means to transport information over long distances.

There has recently been much interest in coupling light with atomic ensembles to develop such a quantum interface. Several proposals have been published to utilize this kind of coupling for spin squeezing \cite{Kuzmich1997PRLv79p4782, Kuzmich1998EPLv42p481, Kuzmich2003IBQuantuminformationwithp231, Thomsen2002PRAv65p061801}, quantum memories \cite{Kozhekin2000PRAv62p033809}, entanglement \cite{Kupriyanov2005PRAv71p032348, Duan2000PRLv85p5643}, quantum teleportation \cite{Duan2000PRLv85p5643, Kuzmich2000PRLv85p5639}, and magnetometry \cite{Geremia2003PRLv91p250801}. Many of these proposals have been realized experimentally using samples of alkali atoms in vapor cells and in magneto-optical traps (MOT) \cite{Hald1999PRLv83p1319, Kuzmich2000PRLv85p1594, Schori2002PRLv89p057903, Julsgaard2001Nv413p400, Julsgaard2004Nv432p482, Geremia2004Sv304p270, Sherson2006Nv443p557}.

The use of alkali metals for the above atom-light interactions has made it possible to access rich hyperfine systems, where the tensor polarizability plays a significant role. There have been several recent studies on the effects of the rank-2 tensor polarizability terms of alkali metals and their application to quantum state preparation and control \cite{DeutschPRA1998v57p1972, Chaudhury2007PRLv99p163002}, atom-light entanglement \cite{Kupriyanov2005PRAv71p032348}, continuous measurement of spin \cite{Geremia2006PRAv73p042112}, and squeezing of an atomic alignment \cite{Cviklinski2007PRAv76p033830}.

In the present paper, we describe the coupling between collective variables of atomic spin and light polarization in a cold $^{87}$Rb ensemble, including effects of multiple hyperfine levels, which manifest as a rank-2 tensor polarizability. We show how the rank-1 or rank-2 terms of the polarizability tensor can be chosen to prevail over the other, leading to Hamiltonians with different symmetries for rotations. As an example, we describe some possible applications of these Hamiltonians, such as measuring atom number, spin squeezing, quantum cloning, and quantum memory, all of which can play an important role in quantum information processing. This analysis is also applicable to other alkali metals, where the polarizability behaves in a similar way, and combinations of the alignment tensor and orientation vector can be used as atomic variables.

This article is organized into four sections. Section~\ref{sec:polar} describes the rubidium system under consideration, as well as the polarizability and Hamiltonian derived from it. In Sec.~\ref{sec:HamEng} we show how different Hamiltonians with different symmetries can be constructed from the tensor polarizability and how these Hamiltonians can be used for several applications. Finally, we present the conclusions in Sec.~\ref{sec:conc}.

%***********************************************************
\section{\label{sec:polar}Polarizability of rubidium atoms}
%***********************************************************

We consider an ensemble of cold $^{87}$Rb atoms interacting with a polarized probe field tuned to the D$_1$ or D$_2$ lines of rubidium and travelling in the $z$-direction. The Hamiltonian describing this interaction can be written as \cite{Happer1972RMPv44p169, Kupriyanov2005PRAv71p032348,
Cohen-Tannoudji1992BAtom-PhotonInteraction}
\begin{equation}\label{eq:polham}
    \hat H_I  = -\sum\limits_{F,F'} {\mathbf{\hat E}^{( - )}  \cdot
    \bm{\hat{\alpha}} _{F,F'} \cdot \mathbf{\hat E}^{( + )}},
\end{equation}
where $\mathbf{\hat E}^{( \pm )}$ is the rotating/counter-rotating term of the probe electric field operator and $\bm{\hat{\alpha}} _{F,F'}$ is the atomic polarizability between a hyperfine ground state $F$ and a hyperfine excited state $F'$.

The atomic polarizability is a rank-2 spherical tensor operator that can be decomposed into three irreducible components
\begin{equation}\label{eq:poltensor}
    \bm{\hat{\alpha}}_{F,F'}  = \bm{\hat{\alpha}}_{F,F'}^{(0)} \oplus
        \bm{\hat{\alpha}}_{F,F'}^{(1)} \oplus
        \bm{\hat{\alpha}}_{F,F'}^{(2)}.
\end{equation}
Hence, the interaction Hamiltonian \eqref{eq:polham} can be written in terms of these irreducible components as
\begin{equation}\label{eq:tensorham}
\begin{split}
    \hat H_I  &= \hat H_I^{(0)}  + \hat H_I^{(1)}  + \hat H_I^{(2)} , \\
    \hat H_I^{(K)}  &= -\sum\limits_{F,F'} {\mathbf{\hat E}^{( - )}
        \cdot {\bm{\hat{\alpha}}_{F,F'}^{(K)} } \cdot \mathbf{\hat
        E}^{( + )}} .
\end{split}
\end{equation}

We will center our attention at the dispersive properties of this interaction, away from any resonances that may cause decoherence through absorption. For the case of rubidium atoms prepared in the $|F=1,\;m=\pm 1\rangle$ ground states, it has been shown \cite{Geremia2006PRAv73p042112,deEchaniz2005JOBv7pS548} that this Hamiltonian can also be expressed in terms of the collective atomic pseudo-spin operators $\mathbf{\hat{J}}$
\begin{equation}
\begin{split}
  \hat J_0  &= \frac{1}{2}\hat{N} , \\
  \hat J_x  &= \frac{1}{2}\sum\limits_k {\left( {\hat F_{x,k}^2 - \hat F_{y,k}^2 } \right)}, \\
  \hat J_y  &= \frac{1}{2}\sum\limits_k {\left( {\hat F_{x,k} \hat F_{y,k}  + \hat F_{y,k} \hat F_{x,k} } \right) } , \\
  \hat J_z  &= \frac{1}{2}\sum\limits_k {\hat F_{z,k} } ,
\end{split}
\end{equation}
where $\hat{N}$ is the atom-number operator, $\hat F_{i,k}$ is the $i$th component of spin operator corresponding to the $k$th atom, and the sum is over all atoms; and the Stokes operators $\mathbf{\hat{S}}$ describing the polarization of the probe field
\begin{equation}
\begin{split}
  \hat S_0  &= \frac{1}{2}\left( {\hat a_ + ^{\dag} \hat a_ +   + \hat a_ - ^{\dag} \hat a_ -  } \right), \\
  \hat S_x  &= \frac{1}{2}\left( {\hat a_ - ^{\dag} \hat a_ +   + \hat a_ + ^{\dag} \hat a_ -  } \right), \\
  \hat S_y  &= \frac{i}{2}\left( {\hat a_ - ^{\dag} \hat a_ +   - \hat a_ + ^{\dag} \hat a_ -  } \right), \\
  \hat S_z  &= \frac{1}{2}\left( {\hat a_ + ^{\dag} \hat a_ +   - \hat a_ - ^{\dag} \hat a_ -  } \right),
\end{split}
\end{equation}
where $\hat a_\pm^{\dag}$ $(\hat a_\pm)$ are the creation (annihilation) operators of the $\sigma^\pm$ modes of the field. The Hamiltonian then takes the form
\begin{subequations}\label{eq:HamPseudo}
\begin{align}
    \hat H_I^{(0)}  &= \frac{4} {3} g\alpha^{(0)}
        \hat S_0 \hat J_0 , \label{eq:HamPseudo-a}\\
    \hat H_I^{(1)}  &= 2g\alpha^{(1)}
        \hat S_z \hat J_z  , \label{eq:HamPseudo-b}\\
    \hat H_I^{(2)}  &= 2g\alpha^{(2)}
        \left( {\hat S_x \hat J_x + \hat S_y \hat J_y + \frac{1}{3}
        \hat S_0 \hat J_0} \right), \label{eq:HamPseudo-c}
\end{align}
\end{subequations}
where $g$ is the form factor of the probe electric field \cite{Scully1997BQuantumOptics}, and
\begin{equation}\label{eq:polff}
\begin{split}
    \alpha^{(0)}  &= (-1)^{2F} \sum\limits_{F'} {\alpha _F^{F'} \left[ {\left( {2F - 1}
        \right)\delta _{F'}^{F - 1}  + \left( {2F + 1} \right)\delta
        _{F'}^F } \right.} \\
        &\quad \left. {+ \left( {2F + 3} \right)\delta _{F'}^{F + 1} }\right], \\
    \alpha^{(1)}  &= (-1)^{2F} \sum\limits_{F'} {\alpha _F^{F'} \left[ { - \frac{{
        2F - 1}} {F}\delta _{F'}^{F - 1}  - \frac{{2F
        + 1}} {{F\left( {F + 1} \right)}}\delta _{F'}^F }\right.} \\
        &\quad \left. {+ \frac{{2F + 3}} {{F + 1}}\delta _{F'}^{F + 1} } \right], \\
    \alpha^{(2)}  &= (-1)^{2F} \sum\limits_{F'} {\alpha _F^{F'} \left[ {\frac{1}
        {F}\delta _{F'}^{F - 1}  - \frac{{2F + 1}}
        {{F\left( {F + 1} \right)}}\delta _{F'}^F }\right. }\\
        &\quad \left. {+ \frac{1} {{F + 1}}\delta _{F'}^{F + 1} } \right], \\
    \alpha _F^{F'} &= \alpha_0 \frac{{\Delta _{F,F'} }}{{\frac{{\Gamma^2
        }}{4} + \Delta _{F,F'}^2 }}\left( { - 1}
        \right)^{J + J' + 2I} \left( {2J' + 1} \right)
        \begin{Bmatrix}
            J' & F' & I \\
            F & J & 1  \\
        \end{Bmatrix}^2 , \\
    \alpha _0  &= \frac{{3\epsilon _0 \hbar \Gamma \lambda^3 }}
        {{8\pi ^2 }},
\end{split}
\end{equation}
with $\delta_{F'}^F$ being the Kronecker delta, $J=1/2\;(J'=1/2,\;3/2)$ the total electronic angular momentum of the ground (excited) state, $I=3/2$ the nuclear spin of the atoms, $\Gamma$ the spontaneous decay rate of the excited state, $\Delta _{F,F'}$ the detuning of the probe from the transition $F \rightarrow F'$, and $\lambda$ the transition wavelength. In Eqs.~\eqref{eq:HamPseudo-a} and \eqref{eq:HamPseudo-c}, the term $\hat S_0 \hat J_0$ represents a global energy shift, and will be ignored for the rest of the paper.

\begin{figure*}
    \includegraphics[width=\columnwidth,keepaspectratio=true]{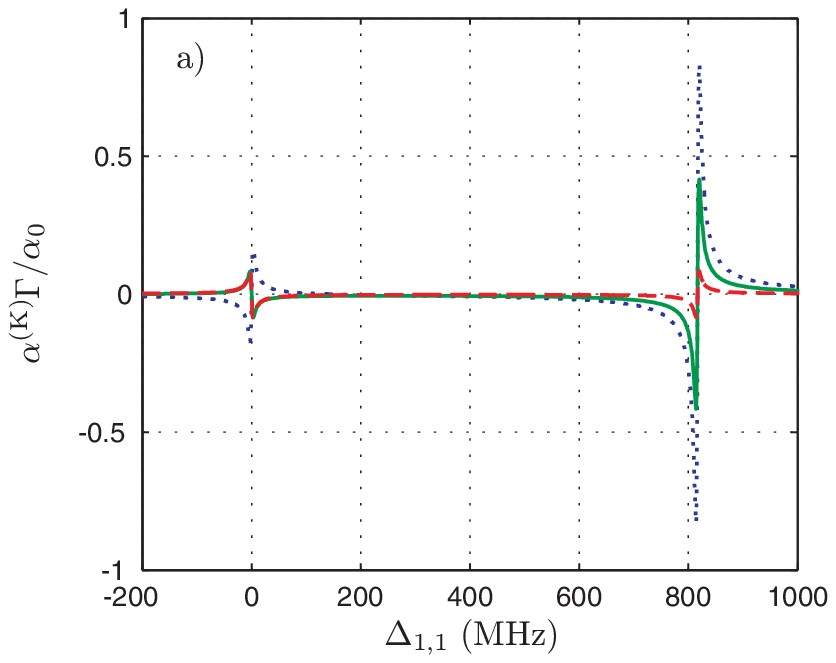}
    \includegraphics[width=\columnwidth,keepaspectratio=true]{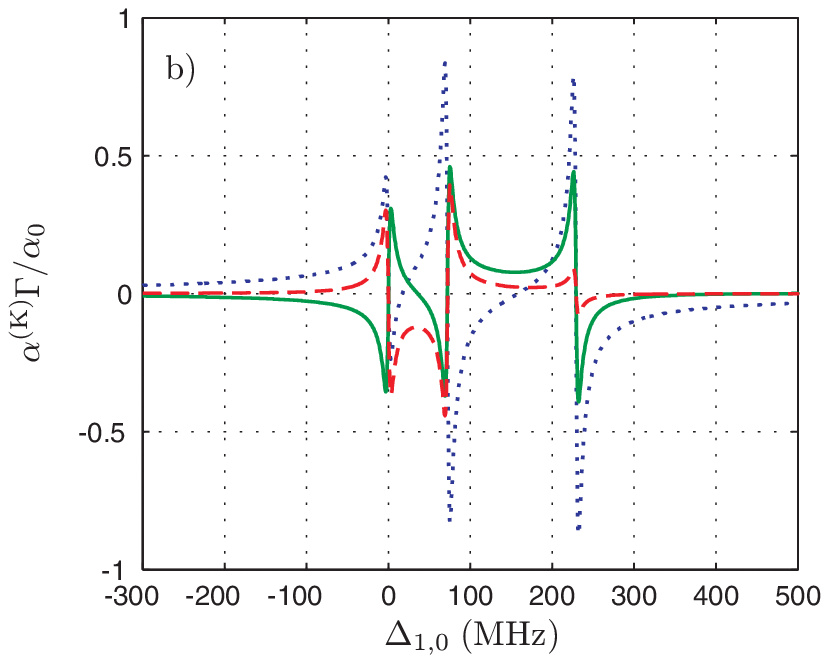}
    \caption{\label{fig:PolarD1-2}Tensor polarizability components of the D$_1$ (a) and D$_2$ (b) lines as a function of probe detuning from the $F=1 \rightarrow F'=1$ (a) and $F=1 \rightarrow F'=0$ (b) transitions. The rank-0 component is shown in dotted blue, the rank-1 component in solid green, and the rank-2 component in dashed red.}
\end{figure*}

Figure \ref{fig:PolarD1-2} shows a numerical calculation of the different tensor polarizability components \eqref{eq:polff} for the D$_1$ (FIG.~\ref{fig:PolarD1-2}a) and D$_2$ (FIG.~\ref{fig:PolarD1-2}b) lines of rubidium as a function of detuning from the $F=1 \rightarrow F'=1$ and $F=1 \rightarrow F'=0$ transitions, respectively. It can be seen in these plots that different ranks of the polarizability are dominant on different ranges of the detuning, particularly on the D$_2$ line. This leads to different rotational symmetries for different values of the detuning, as will be explained in the following Section.

\begin{figure}
    \includegraphics[width=\columnwidth,keepaspectratio=true]{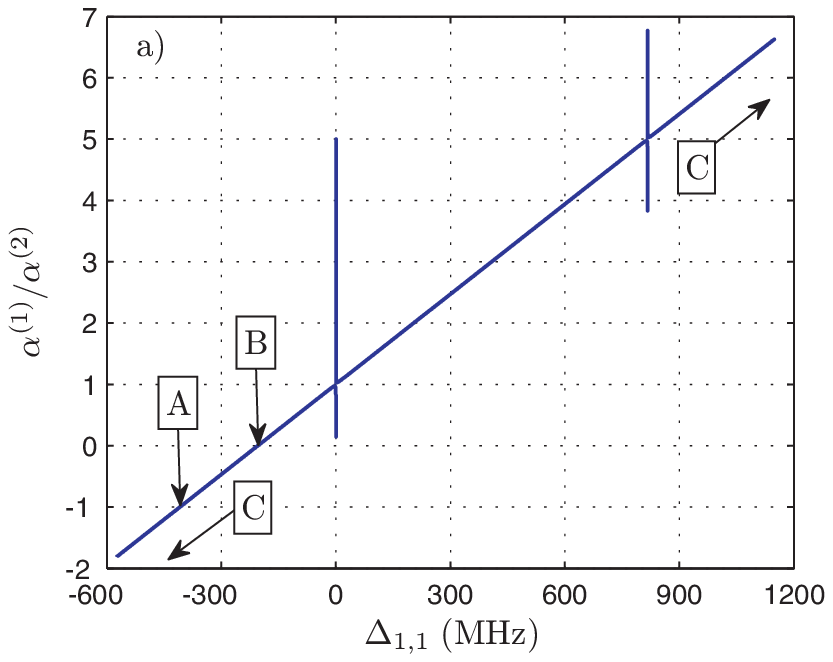}
    \includegraphics[width=\columnwidth,keepaspectratio=true]{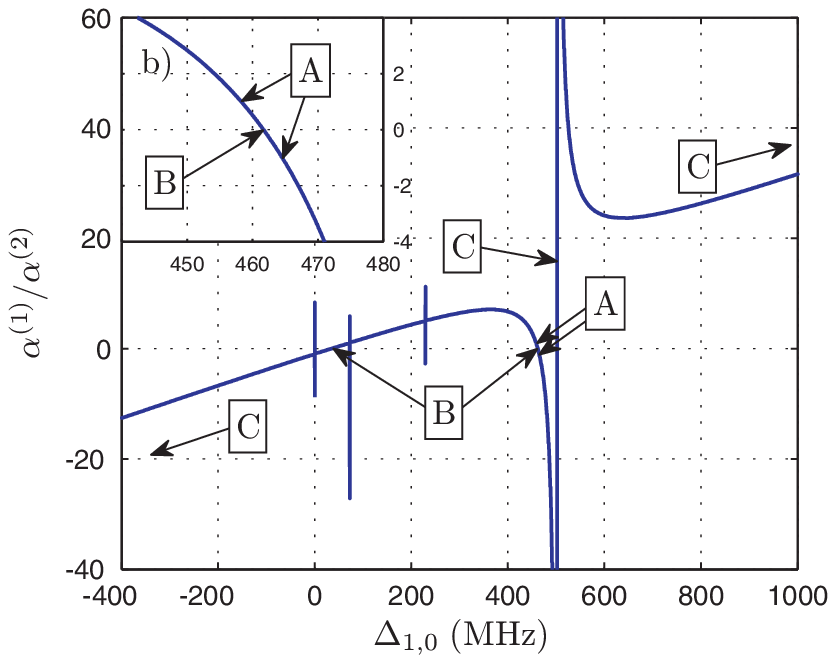}
    \caption{\label{fig:RatioD1-2}Ratio between the rank-1 and rank-2 tensor polarizability components of the D$_1$ (a) and D$_2$ (b) lines as a function of probe detuning from the $F=1 \rightarrow F'=1$ (a) and $F=1 \rightarrow F'=0$ (b) transitions. There are narrow dispersive-like features at the transition detunings which appear as glitches due to the numerical resolution. The labelled arrows A, B and C correspond to points of interest described in the text.}
\end{figure}

It can also be seen (see FIG.~\ref{fig:RatioD1-2}) that the ratio between the rank-1 and rank-2 components of the polarizability grows with the detuning, exhibiting a dispersive-like behavior at the resonant frequencies and on the D$_2$ line at $\Delta_{1,0}=501.7$~MHz, where the rank-2 polarizability crosses zero. The asymptotic rate of growth is $\sim 5$~GHz$^{-1}$ for the D$_1$ line and $\sim 30$~GHz$^{-1}$ for the D$_2$ line. The labelled points A, B and C in FIG.~\ref{fig:RatioD1-2} correspond to values of $\alpha^{(1)}=\pm\alpha^{(2)}$, $\alpha^{(1)}=0$ and $\alpha^{(2)}=0$ respectively, and will be discussed in more detail in the following section.

%**************************************************
\section{\label{sec:HamEng}Hamiltonian engineering}
%**************************************************

The frequency dependence of the polarizability can be used to engineer Hamiltonians \textit{\`{a} la carte}, i.e. by selecting a range of detunings where $\alpha^{(1)}$ or $\alpha^{(2)}$ is predominant, one can choose between the Hamiltonians $\hat{H}_I^{(1)}$ and $\hat{H}_I^{(2)}$ respectively, which exhibit different rotational symmetries.

%-----------------------------------------
\subsection{Symmetries of the Hamiltonian}
As mentioned above, the Hamiltonian of Eq.~\eqref{eq:HamPseudo} may have different symmetries for rotations R depending on the values of $\alpha^{(1)}$ and $\alpha^{(2)}$, where R is an element of the group of rotations generated by $\hat S_i + \hat J_i$. In the case of $\alpha^{(1)} = \alpha^{(2)} = \alpha$, this Hamiltonian will take the form
\begin{equation}
    \hat H_{I}  = 2g\alpha \mathbf{\hat S} \cdot \mathbf{\hat J}, \label{eq:HamJS}
\end{equation}
which is rotationally invariant, and thus has full rotational symmetry. The condition $\alpha^{(1)} = \alpha^{(2)}$ necessary to obtain this Hamiltonian is achieved at a detuning $\Delta_{1,0}=458$~MHz on the D$_2$ line (labelled A in FIG.~\ref{fig:RatioD1-2}b) \footnote{Notice that the condition $\alpha^{(1)} = \alpha^{(2)}$ is also fulfilled on every resonance, but is accompanied by strong absorption.}. It is also possible to obtain $\alpha^{(1)} = -\alpha^{(2)}$ at $\Delta_{1,1}=-408$~MHz on the D$_2$ line, but in this case the Hamiltonian is $\hat{H}_{I} \propto \hat{S}_x \hat{J}_x + \hat{S}_y \hat{J}_y - \hat{S}_z \hat{J}_z$. However, in terms of the angular momentum variables $\mathbf{\bar{J}} \equiv (-\hat{J}_x,\hat{J}_y,-\hat{J}_z)$, $\mathbf{\bar{S}} \equiv (\hat{S}_x,-\hat{S}_y,-\hat{S}_z)$, this has also the form of Eq.~\eqref{eq:HamJS}. Thus, full rotational symmetry also holds for this condition, but for a slightly different description of the spin and polarization variables.

In the case that the rank-1 polarizability vanishes, the Hamiltonian will take the form of Eq.~\eqref{eq:HamPseudo-c}
\begin{equation}
    \hat H_{I} = 2g\alpha^{(2)}
        \left( {\hat S_x \hat J_x + \hat S_y \hat J_y} \right), \label{eq:HamXY}
\end{equation}
which in this case is invariant for rotations about $z$. The condition $\alpha^{(1)} = 0$ is achieved at the detunings labelled B in FIG.~\ref{fig:RatioD1-2}, that is $\Delta_{1,1}=-204$~MHz on the D$_1$ line or $\Delta_{1,0}=462$~MHz on the D$_2$ line \footnote{Also at $\Delta_{1,0}=36$~MHz on the D$_2$ line, but together with strong absorption due to the proximity of the resonances.}. In the case that any of the two terms of this Hamiltonian can be neglected, it is also possible to obtain Hamiltonians which are invariant for rotations about the $x$ or $y$ axes.

Conversely, in the case of $\alpha^{(2)} = 0$, the Hamiltonian will be that of Eq.~\eqref{eq:HamPseudo-b}, which exhibits symmetry for rotations about $z$. A detuning of $\Delta_{1,0}=502$~MHz on the D$_2$ line (labelled C in FIG.~\ref{fig:RatioD1-2}) is needed to fulfill this condition exactly. However, due to the fact that the ratio $\alpha^{(1)}/\alpha^{(2)}$ grows with the detuning, it is also possible to have $\alpha^{(1)} \gg \alpha^{(2)}$ at sufficiently large detunings and approximate the Hamiltonian in \eqref{eq:HamPseudo-b}.

%-----------------------------------------
\subsection{Applications of the Hamiltonian}
These Hamiltonians could be used for numerous applications, some of which we describe in the following subsections. Note that there are several practicable ways to prepare these Hamiltonians, as a consequence of the symmetries explained in the previous subsection. However, for simplicity, here we only describe one of such feasible preparations.

\subsubsection{\label{sec:num}Atom-number measurement}
One of the possible Hamiltonians to prepare is that for atom-number measurement. The pseudo-spin is prepared in its $x$-component, i.e. $\langle\hat{J}_x\rangle = \langle\hat{J}_0\rangle$, $\langle\hat{J}_y\rangle = 0$, $\langle\hat{J}_z\rangle = 0$; and the polarization of the probe in its $z$-component, i.e. $\langle\hat{S}_z\rangle = \langle\hat{S}_0\rangle$, $\langle\hat{S}_x\rangle = 0$, $\langle\hat{S}_y\rangle = 0$. The probe is also tuned where $\alpha^{(1)}=0$ (labelled B in FIG.~\ref{fig:RatioD1-2}) \footnote{Also, there is a range of large $\alpha^{(2)}$ between $F'=0$ and $F'=1$ on the D$_2$ line, but together with high absorption.}, to obtain the effective Hamiltonian of Eq.~\eqref{eq:HamXY}
\begin{equation}
    \hat H_{\text{eff}} = 2g\alpha^{(2)}
        \left( {\hat S_x \hat J_x + \hat S_y \hat J_y} \right).
\end{equation}

In the Heisenberg picture and to first order, this Hamiltonian will rotate $\hat{S}_z$ into $\hat{S}_y$ proportionally to $\hat{J}_x$, which is in turn proportional to the number of atoms. That is
\begin{equation}
\begin{split}
  \hat S_y^{\text{(out)}}  &= - \frac{{2g\tau \alpha ^{(2)} }} {\hbar }\hat J_x^{\text{(in)}} \hat S_z^{\text{(in)}} , \\
  \hat S_z^{\text{(out)}}  &= \hat S_z^{\text{(in)}},
\end{split}
\end{equation}
where we have only kept the non-zero classical terms. Hence, by measuring the rotation into $\hat S_y$, one can measure the number of atoms.

It is important to note that due to the fact that this Hamiltonian is rotationally invariant about $z$, it is possible to prepare the pseudo-spin in any direction on the $xy$ plane and measure the rotation of the polarization into the orthogonal direction.

\subsubsection{\label{sec:squeez}Spin squeezing and atom-light entanglement}
It is also possible to produce the well-studied QND Hamiltonian employed for spin squeezing \cite{Kuzmich1998EPLv42p481, Kuzmich2000PRLv85p1594, Kuzmich2003IBQuantuminformationwithp231, deEchaniz2005JOBv7pS548} and atom-light entanglement \cite{Kupriyanov2005PRAv71p032348}. In this case, $\langle\hat{J}_x\rangle = \langle\hat{J}_0\rangle$ and $\langle\hat{S}_x\rangle = \langle\hat{S}_0\rangle$ will be prepared, and the probe will be tuned to a region of predominant $\alpha^{(1)}$ and vanishing $\alpha^{(2)}$ (labelled C in FIG.~\ref{fig:RatioD1-2}), away from any transitions to avoid scattering. The D$_2$ line is more favorable for this application, as the ratio $\alpha^{(1)}/\alpha^{(2)}$ grows faster with detuning. For a detuning large enough, scattering and the rank-2 polarizability can be neglected, producing an effective Hamiltonian \footnote{For a detuning such that $\alpha^{(2)}=0$ the approximate sign turns into an equal sign.}
\begin{equation}
    \hat H_{\text{eff}}  \approx 2g \alpha^{(1)} \hat S_z \hat J_z .
\end{equation}

To first order, this Hamiltonian will produce the following input/output relations
\begin{equation}\label{eq:InOut}
\begin{split}
    \hat S_y^{\text{(out)}}  &\approx \hat S_y^{\text{(in)}}  +
        \frac{2g\tau}{\hbar}\alpha^{(1)}\hat J_z^{\text{(in)}} \hat S_x^{\text{(in)}} , \\
    \hat S_z^{\text{(out)}}  &= \hat S_z^{\text{(in)}} , \\
    \hat J_y^{\text{(out)}}  &\approx \hat J_y^{\text{(in)}}  +
        \frac{2g\tau}{\hbar}\alpha^{(1)}\hat S_z^{\text{(in)}} \hat J_x^{\text{(in)}} , \\
    \hat J_z^{\text{(out)}}  &= \hat J_z^{\text{(in)}} ,
\end{split}
\end{equation}
where $\hat{J}_z$ is mapped onto $\hat{S}_y$ and $\hat{S}_z$ is mapped onto $\hat{J}_y$, entangling atoms and light \cite{Kupriyanov2005PRAv71p032348}.

Furthermore, since the $z$-components are unchanged owing to the symmetry of the Hamiltonian, this interaction can be used to perform a QND measurement of such components. If the initial spin state was prepared in a minimum uncertainty state (coherent spin state), then this QND measurement will reduce its uncertainty below the projection-noise limit, producing spin squeezing.

\subsubsection{\label{sec:clon}Quantum cloning}
Choosing the initial polarizations $\langle\hat{S}_z\rangle = -\langle\hat{S}_0\rangle$ and $\langle\hat{J}_z\rangle = \langle\hat{J}_0\rangle$, and a detuning where the rank-1 polarizability vanishes (labelled B in FIG.~\ref{fig:RatioD1-2}), it is possible to produce the two-mode squeezing Hamiltonian
\begin{equation}
\begin{split}
    \hat H_{\text{eff}} &= 2g\alpha ^{(2)} \left( {\hat S_x \hat J_x  + \hat S_y \hat J_y }
        \right) \\
    & = \beta \left( { P_L X_A + X_L P_A } \right),
\end{split}
\end{equation}
where $\beta=2g\alpha^{(2)}\sqrt{-\langle\hat{S}_z\rangle \langle\hat{J}_z\rangle}$ and
\begin{equation}
\begin{split}
  \left( {\hat X_L ,\hat P_L } \right) &\equiv {{\left( {\hat S_y ,\hat S_x } \right)}
      \mathord{\left/ {\vphantom {{\left( {\hat S_y ,\hat S_x } \right)} {\sqrt {-\langle\hat{S}_z\rangle } }}} \right. \kern-\nulldelimiterspace} {\sqrt {-\langle\hat{S}_z\rangle } }}, \\
  \left( {\hat X_A ,\hat P_A } \right) &\equiv {{\left( {\hat J_x ,\hat J_y } \right)}
      \mathord{\left/ {\vphantom {{\left( {\hat J_x ,\hat J_y } \right)} {\sqrt {\langle\hat{J}_z\rangle } }}} \right. \kern-\nulldelimiterspace} {\sqrt {\langle\hat{J}_z\rangle } }},
\end{split}
\end{equation}
are the light and atomic canonical variables respectively.

In this configuration, the ensemble acts as a phase-insensitive amplifier for the light, and vice versa. In fact,
\begin{equation}
\begin{split}
  \frac{d}{{dt}}\left( {\hat X_L ,\hat P_L } \right) &= \frac{\beta }
    {\hbar }\left( {X_A ,-P_A } \right), \\
  \frac{d}{{dt}}\left( {\hat X_A ,\hat P_A } \right) &= \frac{\beta }
    {\hbar }\left( {X_L ,-P_L } \right),
\end{split}
\end{equation}
where the light variables are cloned on the atomic ones, and vice versa.

An application in quantum communication is quantum cloning of gaussian states \cite{Braunstein2001PRLv86p4938, Cerf2000PRLv85p1754, Cerf2001IPUniversalcopyingofp11, Cerf2002EPJDv18p211, Cerf2006IBProgressinOpticsp455, Fiurasek2001PRLv86p4942, Fiurasek2004PRLv93p180501}. In that application, the light state is amplified and then split into two (approximate) clones, while the anti-clone remains in the state of the atoms.

\subsubsection{\label{sec:mem}Quantum memory}
In the case that the rank-1 and rank-2 polarizabilities are equal, it is possible to obtain the Hamiltonian of Eq.~\eqref{eq:HamJS}, which will cause the spin and polarization vectors to precess around each other according to
\begin{equation}
\begin{split}
    \frac{d}{{dt}}\mathbf{\hat S} = \frac{2g\alpha}{\hbar} \mathbf{\hat J} \times \mathbf{\hat S}, \\
    \frac{d}{{dt}}\mathbf{\hat J} = \frac{2g\alpha}{\hbar} \mathbf{\hat S} \times \mathbf{\hat J}.
\end{split}
\end{equation}
These equations have the general solution
\begin{widetext}
\begin{equation}
\begin{split}
  \mathbf{\hat J}^{\text{(out)}}  &= \mathbf{\hat{u}}^{\text{(in)}} \left( {\mathbf{\hat{J}}^{\text{(in)}} \cdot \mathbf{\hat{u}}^{\text{(in)}}} \right ) + \mathbf{\hat{V}}^{\text{(in)}} \times \mathbf{\hat{u}}^{\text{(in)}}\cos \left( {\phi} \right) + \mathbf{\hat{V}}^{\text{(in)}} \sin \left( {\phi} \right), \\
  \mathbf{\hat S}^{\text{(out)}}  &= \mathbf{\hat{u}}^{\text{(in)}} \left( {\mathbf{\hat{S}}^{\text{(in)}} \cdot \mathbf{\hat{u}}^{\text{(in)}}} \right ) - \mathbf{\hat{V}}^{\text{(in)}} \times \mathbf{\hat{u}}^{\text{(in)}}\cos \left( {\phi} \right) - \mathbf{\hat{V}}^{\text{(in)}} \sin \left( {\phi} \right),
\end{split}
\end{equation}
\end{widetext}
where
\begin{equation}
\begin{split}
 \mathbf{\hat{u}}^{\text{(in)}} &= \frac{\mathbf{\hat{J}}^{\text{(in)}} + \mathbf{\hat{S}}^{\text{(in)}}}{|\mathbf{\hat{J}}^{\text{(in)}} + \mathbf{\hat{S}}^{\text{(in)}}|}, \\
 \mathbf{\hat{V}}^{\text{(in)}} &= \frac{\mathbf{\hat{J}}^{\text{(in)}} \times \mathbf{\hat{S}}^{\text{(in)}}}{|\mathbf{\hat{J}}^{\text{(in)}} + \mathbf{\hat{S}}^{\text{(in)}}|}, \\
 \phi &= \frac{2g\alpha\tau}{\hbar} |\mathbf{\hat{J}}^{\text{(in)}} + \mathbf{\hat{S}}^{\text{(in)}}|.
\end{split}
\end{equation}

If we choose $|\langle\mathbf{\hat{J}}^{\text{(in)}}\rangle| = |\langle\mathbf{\hat{S}}^{\text{(in)}}\rangle|$, the spin and Stokes vectors will be exchanged after an interaction time $\tau = \pi\hbar/(2g\alpha | \langle\mathbf{\hat{J}}^{\text{(in)}} + \mathbf{\hat{S}}^{\text{(in)}}\rangle |)$, allowing for storage and retrieval of a quantum polarization state into and from the atomic spin.

Another way of realizing a quantum memory is to use a similar scheme to that of cloning described above. In this case, by choosing the initial polarizations $\langle\hat{S}_z\rangle = \langle\hat{S}_0\rangle$ and $\langle\hat{J}_z\rangle = \langle\hat{J}_0\rangle$, and a detuning where the rank-1 polarizability vanishes (labelled B in FIG.~\ref{fig:RatioD1-2}), it is possible to produce the Hamiltonian
\begin{equation}
    \hat H_{\text{eff}} = \eta \left( { X_L X_A + P_L P_A } \right),
\end{equation}
where $\eta=2g\alpha^{(2)}\sqrt{\langle\hat{S}_z\rangle \langle\hat{J}_z\rangle}$ and
\begin{equation}
\begin{split}
  \left( {\hat X_L ,\hat P_L } \right) &\equiv {{\left( {\hat S_x ,\hat S_y } \right)}
      \mathord{\left/ {\vphantom {{\left( {\hat S_x ,\hat S_y } \right)} {\sqrt {\langle\hat{S}_z\rangle } }}} \right. \kern-\nulldelimiterspace} {\sqrt {\langle\hat{S}_z\rangle } }}, \\
  \left( {\hat X_A ,\hat P_A } \right) &\equiv {{\left( {\hat J_x ,\hat J_y } \right)}
      \mathord{\left/ {\vphantom {{\left( {\hat J_x ,\hat J_y } \right)} {\sqrt {\langle\hat{J}_z\rangle } }}} \right. \kern-\nulldelimiterspace} {\sqrt {\langle\hat{J}_z\rangle } }},
\end{split}
\end{equation}
are the light and atomic canonical variables respectively. The equations of motion for these variables will be \cite{Mishina2007PRAv75p042326,Cviklinski2007PRAv76p033830}
\begin{equation}
\begin{split}
  \frac{d}{{dt}}\left( {\hat X_L ,\hat P_L } \right) &= \frac{\eta }
    {\hbar }\left( {P_A ,-X_A } \right), \\
  \frac{d}{{dt}}\left( {\hat X_A ,\hat P_A } \right) &= \frac{\eta }
    {\hbar }\left( {P_L ,-X_L } \right),
\end{split}
\end{equation}
where the light variables are mapped onto the atomic ones and vice versa. The read-out variables in this case will be negated with respect to the ones initially stored. Notice also that only the quantum variables can be stored and not the full vector, like in the previous case.

%**************************************
\section{\label{sec:conc}Conclusions}
%**************************************

We have described the coupling between collective variables of atomic spin and light polarization in a cold $^{87}$Rb ensemble, including effects of multiple hyperfine levels. The polarizability for this kind of system takes the form of a rank-2 tensor. We have shown how the rank-1 or rank-2 terms of the polarizability tensor can be chosen to dominate over the other for different values of the probe detuning on the D$_1$ or D$_2$ lines, the latter showing a richer variety of points of interest. Using this feature of the polarizability, we have engineered Hamiltonians with different symmetries for rotations, which can have applications in schemes for measuring atom number, spin squeezing, quantum cloning, and quantum memory. These applications play an important role in quantum information processing, in particular, for atom-light quantum interfaces.

It is important to note that even though our analysis was centered on the interaction of light with a $^{87}$Rb ensemble, it is still applicable to other alkali metals, where the polarizability behaves in a similar way, and combinations of the alignment tensor and orientation vector can be used as atomic variables.

%**************************************
\begin{acknowledgements}
    This work was funded by the Spanish Ministry of Science and Education under the LACSMY project (Ref. FIS2004-05830) and the Consolider-Ingenio 2010 Project ``QOIT''.
\end{acknowledgements}

%***********************************************************
%\bibliography{//photon/expquantum/Literature/eprints/QOgroup}
\bibliography{SpinPolInt7.bbl}

\end{document}